\documentclass[aps,twocolumn,nofootinbib,groupedaddress]{revtex4-1} 
\usepackage{amssymb,amsmath,amstext,amsfonts}
\usepackage{bbold}
\usepackage{bm}
\usepackage{graphicx}
\usepackage{xcolor}
\usepackage{lipsum}
\usepackage{braket}
\usepackage{comment}
\usepackage{hyperref}

\newcommand{\Mp}{M_{\rm Pl}}
\newcommand{\dd}{\mathrm{d}}
\newcommand{\R}{\zeta}
\newcommand{\F}{\mathcal{F}}
\newcommand{\Nf}{N_\mathrm{flavor}}

\begin{document}

\title{Inflationary flavor oscillations and the cosmic spectroscopy}

\author{Lucas Pinol}
\affiliation{Instituto de F\'isica Te\'orica UAM-CSIC, c/ Nicol\'as Cabrera 13-15, 28049 Madrid, Spain}
\author{Shuntaro Aoki}
\affiliation{Department of Physics, Chung-Ang University, Seoul 06974, Korea}
\author{S\'ebastien Renaux-Petel}
\affiliation{Institut d\char39Astrophysique de Paris, GReCO, UMR 7095 du CNRS et de Sorbonne Universit\'e, 98bis boulevard Arago, 75014 Paris, France}
\author{Masahide Yamaguchi}
\affiliation{Department of Physics, Tokyo Institute of Technology,
Tokyo 152-8551, Japan}

\begin{abstract}
Inflationary scenarios motivated by high-energy physics generically contain a plethora of degrees of freedom beyond the primordial curvature perturbation.
The latter interacts in a simple way with what we name ``inflationary flavor eigenstates'', which differ, in general, from freely propagating ``mass eigenstates''.
We show that the mixing between these misaligned states results in new striking behaviors in the squeezed limit of the curvature perturbation three-point function, depending not only on the mass spectrum but also on the ``mixing angles'' of the theory.
These results bring about a new perspective on the cosmological collider program: contrary to a widespread belief, the primordial signal needs not be dominated by the lightest extra degree of freedom. 
Instead, we show that it may display either modulated oscillations, a broken power law, or a transition from oscillations to a power law, thus offering a detailed \textit{cosmic spectroscopy} of the particle content of inflation.
\vspace{1.4cm}
\end{abstract}

\maketitle

\section{Introduction}

In order to draw conclusions about fundamental physics at the very high energy scales probed during cosmic inflation, an important piece of information is still missing: what is the particle content at play in the primordial universe?
The simplest models of inflation are of the single-clock kind: one scalar field both drives the background evolution and seeds the large-scale inhomogeneities that we observe nowadays in the cosmic microwave background and in the distribution of matter~\cite{Starobinsky:1980te,Guth1981inflation,Albrecht:1982wi,Linde:1983gd,Mukhanov:1985rz,Sasaki:1986hm}.
Although this class of models is phenomenologically viable~\cite{Akrami:2018odb,Planck:2019kim}---provided the scalar potential is sufficiently flat---we expect single-field inflation to be only an approximate emergent description stemming from a more fundamental theory.
Indeed, not only are flat scalar potentials sensitive to Planck-scale physics, but candidate theories at these scales, such as string theory, generically predict a plethora of active particles (see, e.g.,~\cite{Baumann:2014nda}).
Crucially, single-clock models of inflation verify consistency relations, most notably relating the three-point function of the primordial curvature perturbation in the squeezed limit to its power spectrum and corresponding spectral index~\cite{Maldacena:2002vr,Creminelli:2004yq,Cheung:2007sv,Creminelli:2012ed,Senatore_2012,Creminelli:2013cga}.
Not only does this relation not hold in more general scenarios, but recent years have seen the development of the so-called cosmological collider program, aiming at unveiling the inflationary particle content through the robust signatures it leaves in the squeezed bispectrum and other soft limits of higher-order correlation functions (see, e.g.,~\cite{Chen:2009we,Chen:2009zp,Baumann:2011nk,Noumi:2012vr,Assassi:2012zq,Arkani-Hamed:2015bza,Lee:2016vti,Meerburg:2016zdz,Iyer:2017qzw,Arkani-Hamed:2018kmz}).

Due to the widespread belief that only the lightest extra particle is observationally relevant, previous studies have limited themselves to a single extra particle (see~\cite{Chen:2016hrz,Chen:2016uwp,Aoki:2020zbj,lu2021missing} about loop corrections from more fields).
In this work, we explain why this idea is misplaced by considering
inflationary scenarios featuring any number of fluctuating scalar degrees of freedom, that we call ``inflationary flavor eigenstates'', coupled to the observable curvature perturbation through a ``portal field''.
Analogous to the mixing of flavor and mass eigenstates for neutrinos and quarks~\cite{Pontecorvo:1957cp,Maki:1962mu,Cabibbo1963,Kobayashi1973}, the inflationary flavor eigenstates mix with the freely propagating ``inflationary mass eigenstates'' through a mixing matrix to be determined by experiments.
Expressed in the language of an effective field theory for fluctuations~\cite{Creminelli:2006xe,Cheung:2007st}, this generic setup encompasses dominant effects of all explicit multifield models of inflation at lowest order in derivatives~\cite{Pinol:2020kvw}.
We show that it leads to striking many-field observational signatures that cannot be mimicked by simpler models, in particular, for particles with masses of order the Hubble scale, as motivated, e.g., by supersymmetry~\cite{Stewart:1994ts,Baumann:2011nk}.
Depending on the mass spectrum, but also on the ``mixing angles'' of the theory, the squeezed three-point function needs not be dominated by the lightest field and may display either modulated oscillations, a broken power law, or a transition between oscillations and a power law.
This non-trivial ``cosmic spectroscopy'' would, therefore, undoubtedly confirm the presence of multiple primordial degrees of freedom and inform us about the fundamental physics at play in the very early Universe.

\section{flavor and mass bases, \\ mixings and interactions}
\label{sec:flavor-mass-basis}

We consider an inflationary background in which all quantities evolve much more slowly than the scale factor $a(t)$, so that the Hubble scale $H= \dot{a} /a$ (a dot denoting a derivative with respect to cosmic time $t$), the slow-roll parameter $\epsilon = - \dot{H}/H^2$, and other homogeneous functions of time can be considered constant over the time scales relevant for the present study.

\subsection{Flavor basis}

In addition to the usual massless comoving curvature perturbation $\R$, we consider the presence of $\Nf$ interacting fields $\F^\alpha$ that are quadratically coupled to $\zeta$ through the following Lagrangian:
\begin{align}
    \mathcal{L}^{(2)}_\mathrm{flavor}=&\frac{a^3}{2} \left[ \delta_{\alpha\beta} \left(\dot{\F}^\alpha\dot{\F}^\beta - \frac{\partial \F^\alpha \partial \F^\beta}{a^2}  \right)-  M_{\alpha\beta}^2 \F^\alpha \F^\beta  \right. \nonumber \\
    & \left. + 4\sqrt{2\epsilon} \Mp \,  \omega  \delta_{\alpha 1}\F^\alpha \dot{\R} \right] \,,
    \label{eq:flavorLagrangian}
\end{align}
where $\partial$ is the spatial derivative, $M^2$ is a symmetric mass matrix, and $\omega/H$ is the dimensionless coupling quantifying the strength of the quadratic interaction between $\R$ and the new sector. For the purpose of this work, it will not be necessary to spell out concrete models of inflation that realize this setup.
Nonetheless, we stress that this Lagrangian for fluctuations is the one found in nonlinear sigma models of inflation with $\Nf+1$ fields.
In this context, the $\F^\alpha$ are entropic perturbations, $\omega$ is the covariant rate of turn of the multifield background trajectory, and $M^2_{\alpha\beta}$ is related to the inflationary potential and target space geometry~\cite{GrootNibbelink:2001qt,Pinol:2020kvw}.\footnote{In such models, there is an extra interaction in the entropic sector, $a^3 \Omega_{\alpha\beta} \dot{\F}^\alpha\F^\beta$, with $\Omega_{\alpha\beta}$ an antisymmetric matrix representing the rates of turn of the entropic basis along the inflationary trajectory~\cite{GrootNibbelink:2001qt,Pinol:2020kvw}. In the context of this paper, we have checked that treating it perturbatively always leads to next-to-leading order corrections only, and we, therefore, omit it.} We call the extra fields, $\F^\alpha$, flavor eigenstates, since their interactions with the adiabatic sector take a specific form:
only $\F^1$ is directly coupled to $\zeta$, which for this reason we dub the portal field. But crucially, these flavor eigenstates mix due to the \textit{a priori} nondiagonal mass matrix $M^2$,  which calls for a study of the mass eigenstates.

\subsection{Mass basis and mixing angles}

Recalling that $M^2$ is considered constant, we define the mass eigenstates $\sigma^i$ and eigenvalues $m^2_i$ (assumed to be positive for simplicity), $i \in\{1,...,\Nf\}$, and the orthogonal matrix $O^\alpha{}_i$ representing the change of basis between the flavor and mass eigenstates: $\F^\alpha = O^\alpha{}_i \sigma^i$. The quadratic Lagrangian in the mass basis reads:
\begin{align}
\label{eq:L2-mass}
    \mathcal{L}^{(2)}_\mathrm{mass}=&\frac{a^3}{2} \left[ \delta_{ij} \left(\dot{\sigma}^i\dot{\sigma}^j - \frac{\partial \sigma^i \partial \sigma^j}{a^2}  \right)-  \sum_{i=1}^{\Nf} m_{i}^2 \left(\sigma^i \right)^2 \right. \nonumber \\
    & \left.+ 4\sqrt{2\epsilon}  \Mp\,  \omega_i  \sigma^i \dot{\R} \right] \,,
\end{align}
with $\omega_i=\omega  \,O^1{}_{i}$, showing that all mass eigenstates are coupled to $\R$ with strengths that depend on their weights in the portal field since $\F^1 = O^1{}_{i} \sigma^i$. Because $O$ is an orthogonal matrix and represents the mixing of the flavor and mass eigenstates, it may be parametrized with mixing angles, just like the PMNS~\cite{Pontecorvo:1957cp,Maki:1962mu} and CKM~\cite{Cabibbo1963,Kobayashi1973} matrices of the Standard Model of particle physics. But actually, among the inflationary flavor eigenstates, only the portal field $\F^1$ is quadratically coupled to $\zeta$, hence it is enough to consider the fact that
$\sum_i \left(O^1{}_{i}\right)^2=1$, which allows us to write the effective interactions $\omega_i$ in terms of only $\Nf-1$ mixing angles $\theta_{1i}$ with $i \geqslant 2$. For example, if the new sector is a triplet, then $\Nf=3$, and one writes
\begin{equation}
\label{eq:mixing angles}
    O^1{}_{i}=\left[\mathrm{cos}(\theta_{12})\mathrm{cos}(\theta_{13}),\mathrm{sin}(\theta_{12})\mathrm{cos}(\theta_{13}),\mathrm{sin}(\theta_{13})\right]_i \,,
\end{equation}
which can easily be extended to a larger number of flavors.

\subsection{Cubic interactions}

One may distinguish three kinds of cubic interactions, depending on the number of exchanged particles in the tree-level Feynman-like diagrams.
In the mass basis, we consider the following ones: 
\begin{align}
\label{eq:L3}
    \mathcal{L}^{(3)}_{\mathrm{single}} &= - \frac{a^3}{H} \sqrt{2\epsilon} \Mp\, \omega_i \sigma^i \left[ \dot{\R}^2  - \frac{ \left(\partial \R \right)^2}{a^2}\right] \,, \\
    \mathcal{L}^{(3)}_{\mathrm{double}} &= 2 a^3 \epsilon H \Mp^2 R_{ij} \sigma^i \sigma^j \dot{\R} \,, \\ 
     \mathcal{L}^{(3)}_{\mathrm{triple}} &= - \frac{a^3}{6} V_{ijk} \sigma^i \sigma^j \sigma^k \,,
\end{align}
which coincide again with the leading-order interactions of nonlinear sigma models of inflation, provided a suitable identification of the fully symmetric tensors $R_{ij}$ (of mass dimension $-2$) and $V_{ijk}$ (of mass dimension $1$)~\cite{Pinol:2020kvw}. Note also that the coupling $\omega_i \sigma^i$ in $\mathcal{L}^{(3)}_{\mathrm{single}}$ is the same as the one in the quadratic mixing and treats on equal footing
time and spatial derivatives of $\R$. In the language of the effective field theory of inflation~\cite{Creminelli:2006xe,Cheung:2007st}, this fact can be understood as the presence
in the unitary gauge of a  mixing operator with the portal field, $\omega  \F^1 \delta g^{00} \to -  \omega_i \sigma^i \left[ 2 \dot{\pi} + \dot{\pi}^2 - (\partial \pi)^2/a^2 \right]$, where we reintroduced $\pi = - \R / H $ the Goldstone boson of broken time diffeomorphisms.

\section{Free fields, \\power spectrum and bispectrum}
\label{sec:formalism}

\subsection{Interaction picture: free fields and interactions}

In this work, we treat perturbatively the coupling between the portal field and the curvature perturbation, which amounts to choosing as a free Lagrangian the first line of $\mathcal{L}^{(2)}_\mathrm{mass}$ in Eq.~\eqref{eq:L2-mass}, together with $\mathcal{L}^{(2)}_\R = a^3 \epsilon \Mp^2 \left[\dot{\R}^2 - \left(\partial \R\right)^2/ a^2 \right]$.
Note that with this choice, each free field is independent, and independent initial conditions---and, therefore, quantum oscillators---may be drawn.
The corresponding mode functions are
\begin{align}
\label{eq:mode-function}
    \R_k(\tau)=&\frac{H/\Mp}{\sqrt{4 \epsilon k^3}} \left(1+ i k\tau\right) e^{-i k \tau}  \,,  \\
    \sigma^i_k(\tau)=&\,
    e^{i\left(\nu_i-1/2\right)\frac{\pi}{2}} \frac{H (-\tau)^{3/2} \sqrt{\pi}}{2} H_{\nu_i}^{(1)}(- k\tau) \,, \\
    & \text{ with } \,\, \nu_i = \left(9/4- m_i^2 /H^2\right)^{1/2} \,,   \nonumber
\end{align}
where we use conformal time such that $\dd t  =  a \dd \tau$,
with $H^{(1)}_{\nu_i}$ the Hankel function of the first kind with parameter $\nu_i$, and where $\nu_i$ may be purely imaginary if $m_i^2 / H^2 > 9/4$.
These free fields then interact via the quadratic coupling $\mathcal{L}^{(2)}_\mathrm{int} = a^3 2\sqrt{2\epsilon} \Mp \omega_i\sigma^i \dot{\R}$, as well as the cubic interactions.

\subsection{Negligible corrections to the power spectrum}

Doing the explicit calculation, we find that
$\mathcal{L}^{(2)}_\mathrm{int}$ leads to a tree-level correction to the dimensionless power spectrum of $\R$ as
$\mathcal{P}_\R = \mathcal{P}_\R^{(0)}\left[1+\frac{\omega^2}{H^2} \sum_i (O^1{}_i)^2 \mathcal{C}(\nu_i)\right]$
with $\mathcal{P}_\R^{(0)}=H^2/(8\pi^2 \epsilon \Mp^2)$ the usual single-field power spectrum, and $\mathcal{C}(\nu_i)$
a constant that depends on the mass of the exchanged particle~\cite{Chen:2009zp,Chen:2012ge,Pi:2012gf}. The effects from extra particles, consisting here in an unobservable rescaling of the scale-invariant power spectrum, must any way remain negligible, consistently with the perturbative expansion in $\omega/H < 1$.

\subsection{Bispectrum: diagrams and shape function}

\begin{figure*}
    \centering
    \includegraphics[scale=0.35]{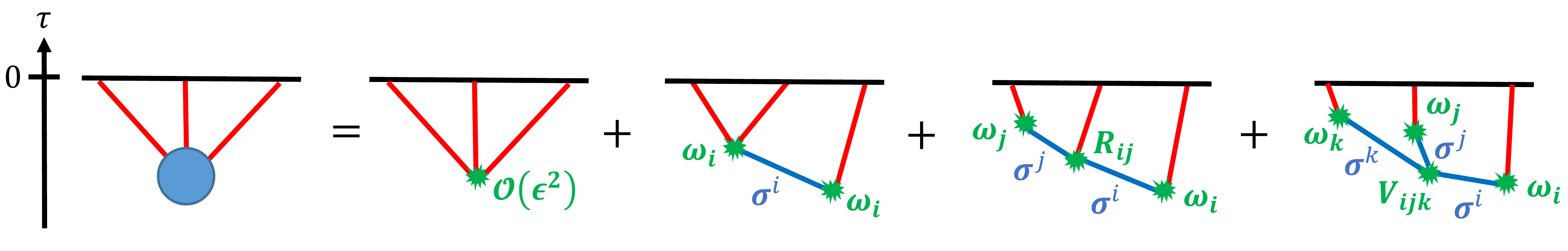}
    \caption{Diagrammatic expression of the tree-level bispectrum of $\R$ (propagators in red), including contributions from the extra scalars $\sigma^i$ (propagators in blue). From left to right: usual single-field slow-roll-suppressed contribution, one-particle $\sigma^i$ exchange with cubic interaction of strength $\omega_i$ (sum over $i)$, two-particles $\sigma^i, \sigma^j$ exchange with cubic interaction of strength $R_{ij}$ (sum over $i,j$), three-particles $\sigma^i, \sigma^j, \sigma^k$ exchange with cubic interaction of strength $V_{ijk}$ (sum over $i,j,k$).}
    \label{fig:diagrams-BS}
\end{figure*}

The crucial novelty of the bispectrum, in contrast to the power spectrum, is that the purely adiabatic contribution from single-field interactions is tiny and suppressed in the squeezed limit once subtle gauge issues are taken into account~\cite{Tanaka:2011aj,Creminelli:2011rh,Pajer:2013ana}.
Therefore, exchanges of extra particles are dominant
in the squeezed limit that, as we show in this paper, is sensitive to all the mass eigenstates.
At leading order, each of the cubic interactions gives an independent contribution to the three-point function, as depicted in the corresponding Feynman-like diagrams of Fig.~\ref{fig:diagrams-BS},
and we will be interested in the associated dimensionless shape function $S$ defined such that $\braket{\hat{\R}_{\vec{k}_1}\hat{\R}_{\vec{k}_2}\hat{\R}_{\vec{k}_3}}=(2\pi)^7 \delta^{(3)}(\vec{k}_1+\vec{k}_2+\vec{k}_3)\left(\mathcal{P}_\R^{(0)}\right)^2 S(k_1,k_2,k_3) / (k_1k_2k_3)^2$.

\section{Squeezed limit:\\ the cosmic spectroscopy}
\label{sec:squezed-limit}

\subsection{Scaling behavior.}
For definiteness, 
we parametrize the wave numbers in the bispectrum as $\left(\vec{k}_1,\vec{k}_2,\vec{k}_3 \right) = \left( \vec{k}_S - \vec{k}_L /2 ,  -\vec{k}_S - \vec{k}_L /2, \vec{k}_L \right)$.
In the so-called squeezed limit with parameter $\kappa= k_L/k_S \ll 1$, we find that the single-exchange bispectrum shape is a sum of individual contributions, with normalized weights $\left(O^1{}_i\right)^2$,
\begin{align}
\label{eq:squeezed-single-bispectrum}
    S_{\mathrm{single}} &\underset{\kappa \ll 1 }{\simeq} \,   - \frac{\pi}{2} \frac{\omega^2}{H^2}  \sum_{i=1}^{\Nf} (O^1{}_i)^2 \mathcal{S}_i \,, \text{\, with  \, } \\
    \mathcal{S}_i &= e^{-\pi \mathrm{Im}(\nu_i)}  \mathrm{Im}\left[\kappa^{1/2+\nu_i} J_+(\nu_i) + \kappa^{1/2-\nu_i} J_-(\nu_i) \right]  \nonumber \,,
\end{align}
with coefficients $J_\pm(\nu_i)$ that depend on the mass of the exchanged particle, see Eq.~\eqref{eq:J-coefficients}.
Note that the mixing angles in $\left(O^1{}_i\right)^2 \leqslant 1$ change the relative weights of each individual contribution, but cannot enhance the overall signal.
For example, if the latter is dominated by a single contribution $(O^1{}_{i_0})^2\mathcal{S}_{i_0}$, then the total signal is still smaller than the corresponding individual one $\mathcal{S}_{i_0}$.
Moreover, this amplitude is small, $(\omega/H)^2 < 1$, a feature that
is not shared by the double- and triple-exchange diagrams, whose amplitudes can naturally be large, a familiar result for the latter~\cite{Chen:2009we,Chen:2009zp}.
However, we chose for this paper to focus on the single-exchange diagram, in order to exemplify the new shape-dependence of the signal in the presence of several additional particles in the simplest technical manner.
We refer the interested reader to the Discussion section for comments about the amplitudes and shapes of the double- and triple-exchange diagrams, see Eq.~\eqref{eq:other-diagrams}, as well as to a separate publication~\cite{Aoki_et_al_inprep} for more explicit results.
Let us now distinguish two physically distinct situations in Eq.~\eqref{eq:squeezed-single-bispectrum}:
\begin{itemize}
    \item For light fields, $m_i < 3H/2$ with real mass parameter $\nu_i$, the $(+)$ and $(-)$ modes are respectively growing and decaying modes as a function of $\kappa$. Therefore, only $J_-(\nu_i)$ is relevant in the above formula that is valid for $\kappa \ll 1$.
    \item For heavy fields, $m_i > 3H/2$ with purely imaginary mass parameter $\nu_i = i \mu_i$, $\mu_i$ being a positive real number, the $(+)$ and $(-)$ modes are oscillating with the same frequency $\mu_i$ in $\mathrm{ln}(\kappa)$ space.
\end{itemize}
In our many-field context, the observable signal is the sum of these individual contributions over all mass eigenstates, and contrary to what may naively be thought, we will show that it may not always be dominated by the lightest field.
But first, we investigate individual contributions, which requires the computation of
\begin{align}
\label{eq:J-coefficients}
    J_\pm(\nu_i) = &\int_0^\infty \dd x_1 \int_{0}^\infty \dd x_2 \,  \mathrm{Im}\left[\left(1 +2ix_1-2x_1^2 \right)  e^{-2ix_1}  \right] \nonumber \\
    &\times A^\pm_{\nu_i}x_1^{-1/2 \pm \nu_i}x_2^{-1/2} e^{-ix_2}    \left(H_{\nu_i}^{(1)}(x_2)\right)^* \,,
\end{align}
where $A^\pm_{\nu_i}$ are $\nu_i$-dependent complex numbers, defined (for $\nu_i \neq 0$) by $H_{\nu_i}^{(1)}(x) \underset{x  \ll 1}{\simeq} A^+_{\nu_i}x^{+\nu_i} + A^-_{\nu_i}x^{-\nu_i}$, and where integrals are regularized in the UV by the usual $i \epsilon$ prescription of the in-in formalism \cite{Weinberg:2005vy}.

\subsection{Investigating the individual contributions}

\begin{figure}
    \centering
    \includegraphics[width=\linewidth]{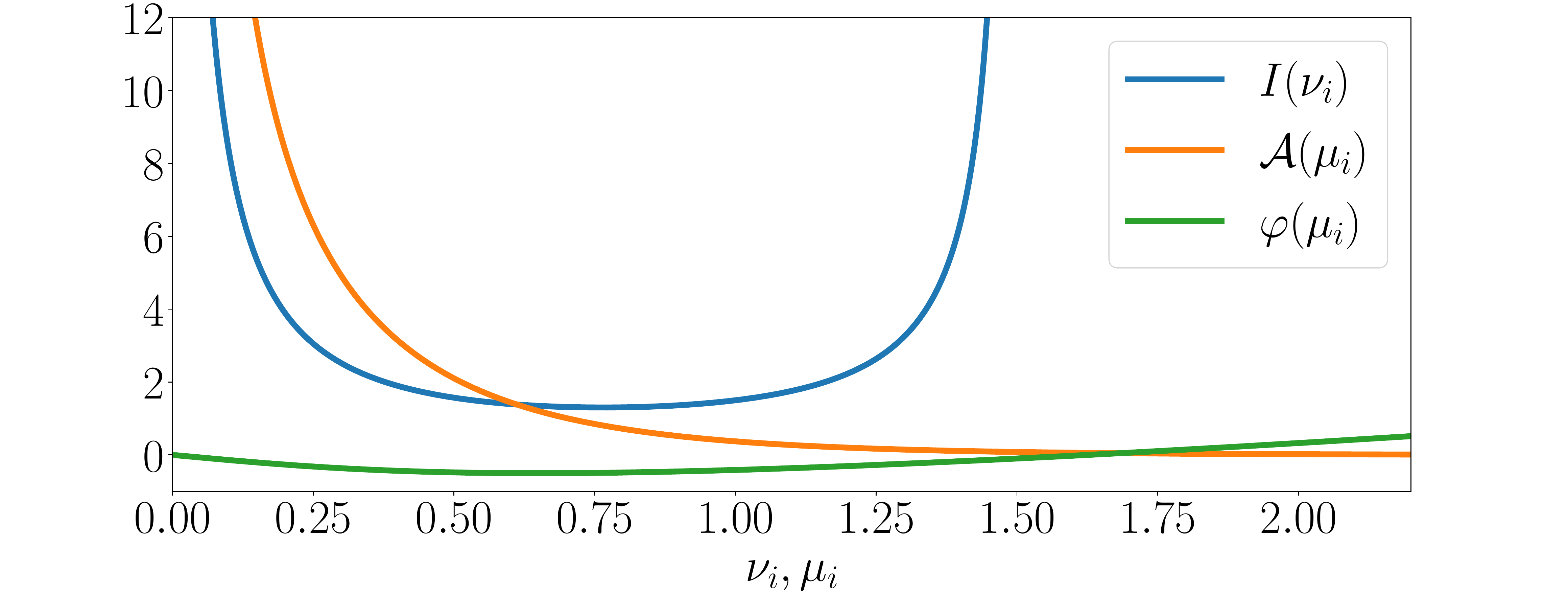}
    \caption{Amplitudes and phase for the different contributions to the squeezed bispectrum. Light fields with parameter $0<\nu_i<3/2$ contribute as a power law proportional to $I(\nu_i)$ (blue line).
    Heavy fields with parameters $\mu_i >0$ contribute with $\mathrm{ln}(\kappa)$ oscillations with an amplitude $\mathcal{A}(\mu_i)$ (orange line) and phase $\varphi(\mu_i)$ (green line).
    We have checked that the apparent divergences of the signal in the $\nu_i, \mu_i \rightarrow 0$ limit are regulated, either by taking into account the other mode $\propto \mathrm{Im}[J_+(\nu_i)]$ in the light case, or by considering the vanishing phase in the heavy case, leading in both situations to the same finite limit, $\kappa^{-1/2}\mathcal{S}_i \rightarrow  -2-15/8[\pi- 4 \mathrm{ln}(2) +\mathrm{ln}(\kappa)] \simeq - 2.69-1.88\,\mathrm{ln}(\kappa)$, and demonstrating the continuity of the signal across the $m_i=3H/2$ threshold.
    The divergence in the $\nu_i \rightarrow 3/2$ limit is a manifestation of the usual burden of IR divergences for massless fields.
    }
    \label{fig:amplitudes}
\end{figure}

Interestingly, it is possible to compute $J_\pm$ analytically, both for light and heavy mass eigenstates.
\begin{itemize}
    \item For light fields and sufficiently squeezed configurations, only $I(\nu_i)=\mathrm{Im}\left[J_-(\nu_i)\right]$ is relevant, and we find
    \begin{align}
    \label{eq:Balpha-light}
        \mathcal{S}_i &= \kappa^{1/2-\nu_i} I(\nu_i) \,, \text{\, with \, }  \\
        I(\nu_i) &= - \frac{2^{-1 + 2 \nu_i} \Gamma(7/2 - \nu_i) \Gamma(\nu_i)\mathrm{cos}(\pi \nu_i)}{\sqrt{\pi} (-1 + 2 \nu_i)[1 + \mathrm{sin}(\pi \nu_i)]} \nonumber \,,
    \end{align}
    where 
    $I(\nu_i)$ is plotted in Fig.~\ref{fig:amplitudes}.
    \item For heavy fields, we rewrite $\mathcal{S}_i$ in order to make explicit the amplitude $\mathcal{A}(\mu_i)$ and the phase $\varphi(\mu_i)$ of the oscillations with frequency $\mu_i$,
    \begin{align}
    \label{eq:Balpha-heavy}
        \mathcal{S}_i=- \kappa^{1/2} & \mathcal{A}(\mu_i) \mathrm{sin}\left[\mu_i \mathrm{ln}(\kappa) - \varphi(\mu_i) \right] \,, \text{\, with \, }  \\
        \mathcal{A}(\mu_i)e^{i\varphi(\mu_i)}=& \frac{\sqrt{\pi } 2^{-1+2 i \mu_i } \tanh (\pi \mu_i ) \Gamma \left(\frac{7}{2}-i \mu_i\right) }{(2\mu_i +i) \Gamma (1-i \mu_i )} \nonumber \\
        \times & e^{-\pi \mu_i}(\coth (\pi  \mu_i )+i\text{csch}(\pi  \mu_i )+1)^2\,, \nonumber
    \end{align}
    and $\mathcal{A}(\mu_i)\,,\varphi(\mu_i)$ are
    plotted in Fig.~\ref{fig:amplitudes}.
\end{itemize}

We have checked that our results reduce exactly to known ones in the limit of a single extra mass eigenstate $\sigma$, both in the light case~\cite{Noumi:2012vr} and in the heavy one~\cite{Arkani-Hamed:2015bza}, even if computed with a completely different method for the latter.

To simplify the discussion about observational signatures, we restrict ourselves in the next paragraphs to the case of two extra scalars, i.e., $\Nf=2$, and therefore a single mixing angle $\theta_{12}$: $O^1{}_{i}=\left[\mathrm{cos}(\theta_{12}),\mathrm{sin}(\theta_{12})\right]_i$, where here one can actually choose $\theta_{12} \in [0,\pi/2]$.
Moreover, we consider masses close to the Hubble scale.
For definiteness, we also choose $m_1^2 > m_2^2$.
Therefore, in the limit $\theta_{12} \rightarrow 0$ (respectively, $\theta_{12} \rightarrow \pi/2$) one recovers the signal from the heavier (respectively, the lighter) field alone.
However, generic values, such as $\theta_{12} = \pi/4$ for which the portal field has equal weights in both mass eigenstates, $\F^1 = (\sigma_1 + \sigma_2)/\sqrt{2}$, result in a new striking phenomenology that we now discuss.

\subsection{Modulated oscillations}

\begin{figure}
    \centering
    \includegraphics[width=\linewidth]{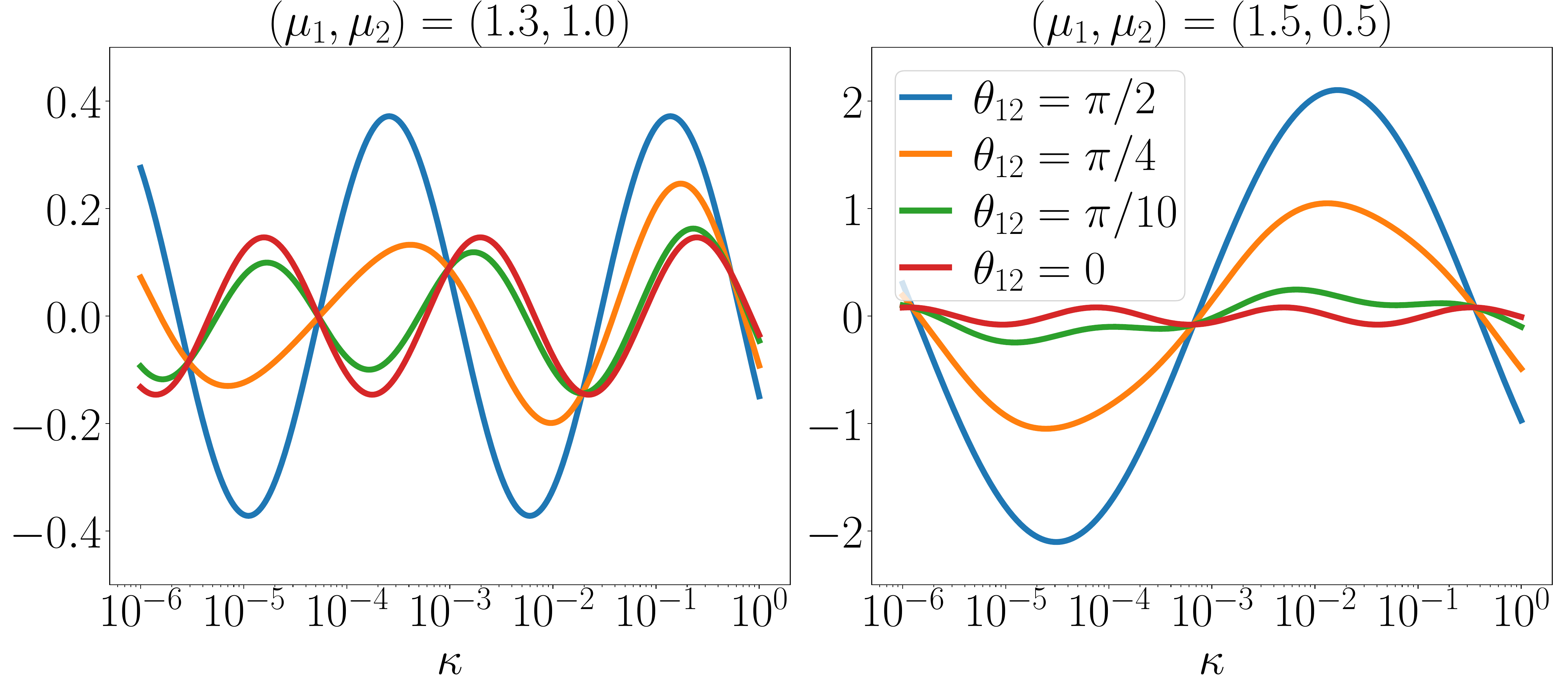}
    \caption{Rescaled signal in the squeezed limit of the bispectrum, $\kappa^{-1/2}\sum_i (O^1{}_i)^2 \mathcal{S}_i$, for two extra heavy fields.
    The left panel corresponds to $(\mu_1,\mu_2) =(1.3,1.0)$ and the right one to $(\mu_1,\mu_2) =(1.5,0.5)$, for different values of the mixing angle $\theta_{12} \in [0,\pi/2]$.
    The modulation of the oscillations from the heavier field is clearly visible, even if only for a small nonzero mixing angle $\theta_{12}$ when the mass hierarchy is too strong.}
    \label{fig:two-heavy-fields}
\end{figure}

We first consider the situation of two heavy fields with both $\nu_{1,2}=i\mu_{1,2}$ purely imaginary, and $\mu_1 > \mu_2 > 0$.
The overall signal is then the sum of two oscillating contributions with different frequencies $\mu_1, \mu_2$ and respective amplitudes and phases, meaning modulated oscillations with frequencies $(\mu_1 \pm \mu_2)/2$.
In Fig.~\ref{fig:two-heavy-fields}, we show the resulting signal for close masses $(\mu_1,\mu_2) =(1.3,1.0)$ (left panel) or differing by an order one factor $(\mu_1,\mu_2)=(1.5,0.5)$ (right panel), for four values of the mixing angle, $\theta_{12} \in \{0, \pi/10, \pi/4,\pi/2 \}$.
While the modulated oscillations are clearly visible for a generic mixing in the case of close masses (see the orange line for $\theta_{12}=\pi/4$ in the left panel), the observation of such an effect for a larger hierarchy of masses requires a smaller mixing angle, and, therefore the portal field to be dominated by the heavier mass eigenstate in order to compensate for its power suppression (see the green line for $\theta_{12}=\pi/10$ in the right panel).
Observing modulated oscillations in the squeezed limit of the bispectrum would be a striking signature of multiple heavy degrees of freedom beyond the inflaton in the very early Universe.
\begin{figure}
    \centering
    \includegraphics[width=\linewidth]{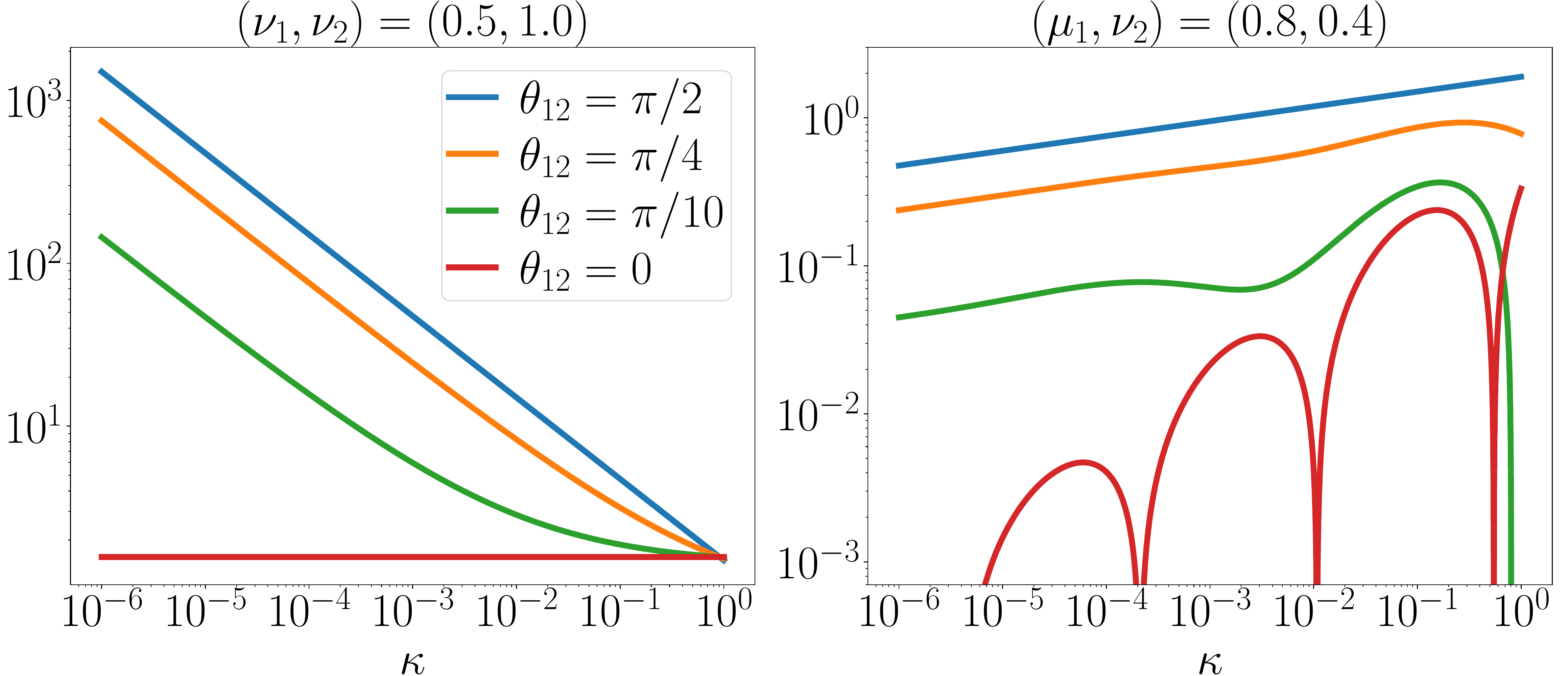}
    \caption{Overall signal $\sum_i (O^1{}_i)^2 \mathcal{S}_i$ in the squeezed limit of the bispectrum, for two extra light fields with $(\nu_1,\nu_2)=(0.5,1.0)$ (left panel), as well as one heavy field and one light one with $(\mu_1,\nu_2)=(0.8,0.4)$ (right panel), for different values of the mixing angle $\theta_{12} \in [0,\pi/2]$.
    The effect of the heavier field is a transition at intermediate squeezing values, between either two different power laws  or an oscillatory signal and a power law, but is only visible for a small nonzero mixing angle $\theta_{12}$.
    }
    \label{fig:light-mixed-fields}
\end{figure}

\subsection{Broken power law}

Another case consists in having only light fields, with real mass parameters $\nu_{1,2}$ and $0 < \nu_1 < \nu_2 < 3/2 $, in which case the resulting signal is the sum of two power laws.
In contrast to the oscillations of heavy fields that periodically vanish, the individual contribution from the lighter field is, in this case, always dominant.
Therefore, it is only if the portal field is dominated by the heavier mass eigenstate, i.e., for small mixing angles, that the squeezed limit of the bispectrum displays a many-field behavior.
Indeed in that case, the overall signal transitions between two different power laws for intermediate values of the squeezing parameter,
as can be seen in the left panel of Fig.~\ref{fig:light-mixed-fields} (in particular, see the green line for $\theta_{12}=\pi/10$).

\subsection{Mixed case}

An interesting possibility is to have a light field and a heavy one in the new sector, with mass parameters $\nu_1=i\mu_1$ purely imaginary and $\nu_2$ real.
Like the previous two-light-fields case, it is only for small mixing angles that the effect of the heavier of the two fields can be seen in the squeezed bispectrum.
In that case, there is a qualitative change of behavior from a certain value of the squeezing parameter, transitioning from oscillations to a power law at intermediate squeezing values,
as can be seen in the right panel of Fig.~\ref{fig:light-mixed-fields} (in particular, see the green line for $\theta_{12}=\pi/10$).

\section{Discussion}
\label{sec:discussion}

In this paper, we have investigated how the mixing of inflationary flavor and mass eigenstates affects the bispectrum of the primordial curvature perturbation in the squeezed limit.
We have identified that cosmological observations are not only sensitive to the mass spectrum of the freely propagating degrees of freedom, but also to mixing angles that measure the weights of the different mass eigenstates in the portal field, which is the only flavor directly coupled to $\zeta$.
Indeed, large-scale fluctuations of these fields of different masses interfere with unequal amplitudes and phases with fluctuations of $\zeta$ on smaller scales when the latter exit the horizon, imprinting a nontrivial pattern in the primordial bispectrum.
Recovering \textit{en passant} several known results of the literature when there is only a single extra degree of freedom, we have shown that single-exchange diagrams display new striking observational signatures: modulated oscillations, a broken power law, or a transition between oscillations and a power law.

Importantly, our results show that the general belief that only the lightest field is relevant in a many-field situation is wrong in the case of heavy fields with $m_i > 3H/2$, and although true in the asymptotically squeezed limit $\kappa \rightarrow 0$ in the case of light fields with $m_i < 3H/2$, and in the mixed case with both light and heavy fields, it is also misplaced for reasonable squeezing values of observational relevance.

The overall amplitude of the signal from the single-exchange diagram considered in this paper is small:  $f_\mathrm{NL}^\mathrm{sq} \sim (\omega/H)^2 < 1$, where $\omega/H$ is the dimensionless coupling between the curvature perturbation and the portal field.
However, this needs not be the case for the  double- and triple-exchange diagrams, with dimensionless couplings $R_{ij}H^2$ and $V_{ijk}/H$ independent from $\omega/H$, and that feature boosted amplitudes (note the negative powers of the scalar power spectrum),

\begin{widetext}
\begin{align}
\label{eq:other-diagrams}
    S_{\mathrm{double}} &\underset{\kappa \ll 1 }{\simeq} \, \frac{1}{4} \frac{\omega^2}{H^2} \frac{1}{\mathcal{P}_\R^{(0)}} \sum_{i,j}  R_{ij} H^2  O^1{}_i O^1{}_j  e^{-\pi \mathrm{Im}(\nu_i+\nu_j)}  \mathrm{Im}\left[\kappa^{1/2+\nu_i} J_+^{\mathrm{double}}(\nu_i,\nu_j) + \kappa^{1/2-\nu_i} J_-^{\mathrm{double}}(\nu_i,\nu_j) \right] \,, \\
    S_{\mathrm{triple}} &\underset{\kappa \ll 1 }{\simeq} \, \frac{\pi^2}{4} \frac{\omega^3}{H^3} \frac{1}{\sqrt{\mathcal{P}_\R^{(0)}}} \sum_{i,j,k}  \frac{V_{ijk}}{H}   O^1{}_i O^1{}_j  O^1{}_k e^{-\pi \mathrm{Im}(\nu_i+\nu_j+\nu_k)}  \mathrm{Im}\left[\kappa^{1/2+\nu_i} J_+^{\mathrm{triple}}(\nu_i,\nu_j,\nu_k) + \kappa^{1/2-\nu_i} J_-^{\mathrm{triple}}(\nu_i,\nu_j,\nu_k) \right] \,, \nonumber
\end{align}
\end{widetext}
with coefficients $J_\pm^{\mathrm{double},\mathrm{triple}}$ that depend
on the masses of the exchanged particles. 
We will report on the detailed behaviors of the double- and triple-exchange diagrams, qualitatively similar to the one presented in this work, in a separate publication~\cite{Aoki_et_al_inprep}.

Our results open several avenues for new research.
Theoretically, it would be interesting to include spinning particles, to consider higher-order correlation functions, and to estimate subleading contributions in not-so-squeezed configurations.
Observationally, it motivates studying how, and to which extent, the various cosmological
probes can fulfill the promise of the primordial cosmic spectroscopy presented here.


\begin{acknowledgments}

We wish to thank Matteo Fasiello, Jacopo Fumagalli, Sadra Jazayeri, Toshifumi Noumi and Denis Werth for useful discussions on the topic of this paper.
L.P. would like to acknowledge support from the “Atracción de Talento” grant 2019-T1/TIC15784.
The work of SA is supported in part by Basic Science Research Program through the National Research Foundation of Korea (NRF) funded by the Ministry of Education, Science and Technology (NRF-2019R1A2C2003738). S.RP is supported by the European Research Council under the European Union’s Horizon 2020 research and innovation programme (grant agreement No 758792, project GEODESI). MY is supported in part by JSPS Grant-in-Aid for Scientific Research Numbers JP18K18764, JP21H01080, JP21H00069. This article is distributed under the Creative Commons Attribution International Licence (\href{https://creativecommons.org/licenses/by/4.0/}{CC-BY 4.0})

\end{acknowledgments}



%

\end{document}